\documentclass[12pt]{article}

\makeatletter%
\def\nottoobig#1{{\hbox{$\left#1\vcenter to1.111\ht\strutbox{}\right.\n@space$}}}
\makeatother%

\makeatletter%
\def\@begintheorem#1#2{\trivlist\item[\hskip\labelsep{\bf #1\ #2}]}
\makeatother
\makeatletter 

\newlength{\filength}
\newsavebox{\gcbox}
\sbox{\gcbox}{\framebox[\filength]{\rule{0ex}{2ex}}}

\newlength{\leftjustindent}
\newlength{\@leftjustindent}
\def\leftjust{\let\\\@leftjustcr\let\end\@endleftjust
  \addtolength{\@leftjustindent}{\leftjustindent}
  \vcenter\bgroup
  \halign\bgroup
    \hbox to\displaywidth{
      \rule{\@leftjustindent}{0ex}$\displaystyle##$\hfill
      }\crcr
}
\def\endleftjust{\crcr\egroup\egroup\endgroup}
\def\@endleftjust#1{\crcr\egroup\egroup\@checkend{#1}\endgroup}
\def\@leftjustcr{\crcr}

\newcommand{\singlespacing}{\let\CS=
\@currsize\renewcommand{\baselinestretch}{1}\tiny\CS}
\newcommand{\singlespacingplus}{\let\CS=
\@currsize\renewcommand{\baselinestretch}{1.25}\tiny\CS}
\newcommand{\doublespacing}{\let\CS=
\@currsize\renewcommand{\baselinestretch}{1.75}\tiny\CS}
\newcommand{\draftspacing}{\let\CS=
\@currsize\renewcommand{\baselinestretch}{2.0}\tiny\CS}

\makeatother

\makeatletter
\clubpenalty=\@highpenalty
\widowpenalty=\@highpenalty
\makeatother

\makeatletter
\newcommand{\niceonespacing}{\let\CS=\@currsize\renewcommand{\baselinestretch}{1.1}\tiny\CS}\newcommand{\nicetwospacing}{\let\CS=\@currsize\renewcommand{\baselinestretch}{1.2}\tiny\CS}
\newcommand{\nicethreespacing}{\let\CS=\@currsize\renewcommand{\baselinestretch}{1.3}\tiny\CS}
\newcommand{\singlespacingplusplus}{\let\CS=\@currsize\renewcommand{\baselinestretch}{1.35}\tiny\CS}
\newcommand{\nicefivespacing}{\let\CS=\@currsize\renewcommand{\baselinestretch}{1.5}\tiny\CS}
\newcommand{\nicesixspacing}{\let\CS=\@currsize\renewcommand{\baselinestretch}{1.6}\tiny\CS}
\makeatother

\makeatletter
\def\@cite#1#2{[#1\if@tempswa , #2\fi]}
\makeatother

\makeatletter
\def\@citex[#1]#2{\if@filesw\immediate\write\@auxout{\string\citation{#2}}\fi
  \def\@citea{}\@cite{\@for\@citeb:=#2\do
    {\@citea\def\@citea{,\linebreak[0]}\@ifundefined
       {b@\@citeb}{{\bf ?}\@warning
       {Citation `\@citeb' on page \thepage \space undefined}}%
\hbox{\csname b@\@citeb\endcsname}}}{#1}}
\makeatother

\begin{document}

\title{
Take-Home Complexity\thanks{
Supported in part 
by grants
NSF-CCR-9322513
and 
NSF-INT-9815095/\protect\linebreak[0]DAAD-315-PPP-g\"u-ab.  
Updated version of 
SIGACT News Complexity
Theory Column~20.}}
\author{Lane A. Hemaspaandra
\\
Department of Computer Science\\
University of Rochester\\
Rochester, NY 14627\\
{\tt  lane@cs.rochester.edu}}
\date{January 20, 2000 }

{\singlespacing

\singlespacing\maketitle

}

\singlespacing

\section{Introduction}

Let us consider the first-year graduate complexity theory course.
Here at the University of Rochester we use Papadimitriou's
book~\cite{pap:b:complexity} and Bovet and Crescenzi's
book~\cite{bov-cre:b:complexity} as the texts, but many
other fine options exist, e.g.,
\cite{sav:b:models,sip:b:theory-of-computation,bal-dia-gab:b:sctI-2nd-ed,bal-dia-gab:b:sctII},
and the author hopes that even more options will soon become 
available~\cite{hem-ogi:b:ccc}.
By ``the first-year graduate complexity theory course,'' 
let us mean a one-semester
course designed for first-year graduate students (not just theory
people, but people from all areas), and probably cross-listed so 
advanced undergraduates can take it.

One lofty goal that such courses often have is to give students a feel
for what it is like to be a complexity theorist.  That is, such
courses, in addition to conveying a certain body of information, often
have the far more difficult goal of giving students a taste of the way
theorists think and work, ideally by having students think and
work in the same way (within the context of the course---but in the
best of cases, even the non-theory students may let their future
research be influenced and informed by a theoretician's love of crisp
formalization, skeptical thought, and lovely theorems/proofs).

Homework, teaching style, exams, and in-class workshops each can all
contribute to this goal.  For example, one can and probably should
make the class sessions very interactive---happily chasing down (and
trying, live-on-the-spot, to explore and prove things about) whatever
directions students ask about (``teacher, teacher, what if you remove the
injectivity requirement from that theorem's statement?''...  ``that's
a wonderful, natural question... let's jump in and see if we can
re-establish a complete characterization... do you think the previous
characterization's proof will still be ok under the alteration you have
proposed?''...)

However, we've found that the very best way of giving students a taste
of the life of a complexity theorist is by including a one-~or
two-week project in the course.  We've tried two quite different types
of projects: research projects and critical-thinking projects.  This
column describes a sample or two of each, and reports on the results
of using the various projects.

\section{Research Projects}
Any researcher can think of countless research projects on which to
send students.  Of course, most are probably overly difficult---at
least relative to the background students have after a semester of
graduate complexity theory.  So the trick here is to make the project
feasible while allowing room for creativity and tying in with a
subject that the course covered in some detail.

In our first-year graduate complexity course here at Rochester, we
cover certain core things each year (namely, the nice, standard stuff
from Papadimitriou and Bovet-Crescenzi), and usually also have time
for a dealer's choice unit.  For example, in one recent year, this
unit was on Sch\"oning's theory of robust algorithms, which led to
more recent work (explicitly or implicitly) on probabilistic and
unambiguous variants of the theory, including connections
to interactive proof
systems and a variety of other topics.
(The paper that started this area 
was~\cite{sch:j:robustness} and since there have been too many 
papers to completely list here, including, for example,
the following papers on the topic and its many 
cousins
\cite{ko:j:helping,bal:t:smart,sch:c:robustness-mfcs-survey,sch:b:complexity-interaction,bal:j:solutions,har-hem:j:rob,vys:unpub:helpers,yam:unpub:helpers,cai-hem-vys:b:promise,for-rom-sip:j:multi-prover,ogi:j:helping,arv-koe-sch:j:helping}.)

Building on that unit, the take-home (week-long) project in that year
was to define and study exponential-time analogs of the standard
(polynomial-time) theory, and to write one's results up in a paper
having the form of a journal submission (actual journal submission not
required!).  This project---about which we were (and remain) least
enthusiastic among those discussed in this column---was the most
successful.  Students handed in nice papers, had (in some cases)
side-stepped some problems one can have on this if one is not careful,
had (in some cases) thought carefully about definitions and models,
and had (in some cases) proved some very nice theorems and outlined
some remaining interesting open issues.

\section{Critical-Thinking Projects}
Research projects, such as the one described in the previous section,
let students simulate the research part of being an theoretical
computer scientist.  Another important component of being a researcher
is the ability to read---skeptically and carefully---technical papers.
Refereeing is an obvious way one does this, but in fact every time one
reads a paper this comes into play.

As critical-thinking (take-home) projects, we like to assign students
some (slightly or enormously) flawed paper on an interesting topic.
Each student is asked to view the paper as a journal submission that
he or she has been asked to referee within a week (hey, it is just a
simulation!,~not reality), and to (within a week) write a detailed
referee report.  

The trick here, in choosing topics, is to find ones that encourage
careful, focused thinking on each student's part---not topics where he
or she can find the answer by hunting up an erratum/corrigendum.  
Two examples that we've used at my school follow.

This year, we gave our first-year complexity course's students a
beautiful paper by Hartmanis and Yesha, ``Computation Times of {NP}
Sets of Different Densities''~\cite{har-yes:j:computation}, and asked
them to referee it.  This paper is a wonder.  It tightly ties
fundamental questions in complexity theory 
($\rm P = PSPACE$?,\ $\rm P=NP$?)\ to 
(in a certain formal sense) issues of whether mathematical
creativity is within the reach of computing
machines~\cite[Section~3]{har-yes:j:computation}.  Assigning this
paper gives the students a chance to read one of the most delightful
and underappreciated papers in our field---and, to boot, one that
strongly motivates the study of computational complexity.  As to the
referee report project on this paper, there is something easy for the
students to find.  One theorem, which is of the form FOO iff BAR, has
a proof that goes like this:
\begin{quote}
  FOO implies BAR because Ni, Ping, and Nee-womm......
  Conversely, if BAR
  does not hold then FOO does not hold because 
  Ekky-ekky-ekky-ekky-z'Bang, zoom-Boing,
  z'nourrrwringnmmm......~QED
\end{quote}
The worry here is that the proof proves the ``only if'' direction
twice (via proving it and then its contrapositive, rather than its
converse), and the ``if'' direction not at all.  This flaw in
the logical argumentation is certainly one that
the course's students should
find and, beyond that, the missing direction in fact does hold and is
sufficiently straightforward that the 
course's students should be able to
provide a correct proof as part of their reports (in reality, these
``should''s proved a bit too optimistic).  In summary, the goals of this
project were to let students learn the beautiful work of the
Hartmanis/Yesha paper, and to give students a chance to exercise and
develop the critical theory-reading skills used daily by theoretical
computer scientists.

Our second example of an interesting critical-thinking take-home
project is quite different.  This one, which we used a year or two ago,
simply asked the students to referee a draft from the early 1980s
entitled ``On the Complexity of Uniqueness Problems,'' by Edwards and
Welsh~(\cite{edw-wel:unpub:unique}, see the discussion
in the paper~\cite{fis-hem-tor:b:witness-iso}, which 
is in part motiviated by the draft of Edwards and Welsh).  
The draft has circulated widely
and has influenced many people---but as far as we know never was even
made into a technical report (probably due to the problem discussed
below).  This bold paper pretty much claims to disprove the
Berman-Hartmanis Isomorphism Conjecture.\footnote{Informally put,
the Berman-Hartmanis Isomorphism Conjecture 
states that
there is essentially just {\em one\/} NP-complete set---that
dresses itself
up in a variety of ways via trivial renaming. 
More formally, the Berman-Hartmanis Isomorphism Conjecture 
says that every two NP-complete sets are polynomial-time
isomorphic.}
To achieve their disproof, the authors
argue that 
Berman-Hartmanis~\cite{ber-har:j:iso} 
were actually conjecturing more than Berman-Hartmanis
said, namely, that Berman-Hartmanis actually intended to make an
extraordinarily strong claim about parsimonious interreducibility of
NP-complete sets.

Unfortunately, the strong claim is {\em so\/} strong that it can
easily be falsified, which is exactly what Edwards and Welsh then do.
Though certainly Berman-Hartmanis neither made nor intended to make
such a strong claim, the direction of---and general intution
behind---the strong claim is very nice and interesting.  In fact, it
is at least plausible to conjecture that NP-complete sets may have even
more in common than mere isomorphism (and, indeed, many natural sets
{\em do\/} have more in common), and many papers 
have
wondered just what that ``more'' may be.  One could say that this
general type of intuition has led people to such notions as
structure-preserving 
reductions~\cite{lyn-lip:j:structure-preserving,aus-atr-pro:j:structure},
witness-isomorphic reductions~\cite{fis-hem-tor:b:witness-iso},
universal
relations~\cite{agr-bis:cOUTbyJOUR:universal,bis:unpub:conp-universality,buh-kad-thi:j:functions},
and much more.

The students' results on this one were mixed.  Almost all picked up
that something was very strange, but their referee reports varied as
to the clarity with which they pinpointed the problem.  In some sense,
this is a more demanding assignment than the Hartmanis-Yesha one, as
the problem is not one of a pure error in logical flow, but rather is
that here one must argue against the authors' opinion about what was
in the minds of Berman and Hartmanis---and what is actually the most
natural expansive version of Berman and Hartmanis's insightful
conjecture (which, to this very day, remains open, though dozens of
papers have been written on the topic; a nice survey by Kurtz,
Mahaney, and Royer covers the progress up to about nine years
ago~\cite{kur-mah-roy:b:iso}, and the study of
isomorphism results has also been quite active
in the years since that survey, see, 
for 
example,~\cite{fen-for-kur-li:c:oracle-toolkit,rog:j:iso-one-way,agr-all-rud:j:reductions-isomorphism}).

{\samepage
\begin{center}
{\bf Acknowledgments}
\end{center}
\nopagebreak \indent I am very grateful to Eric Allender, Someth
Biswas, and Lance Fortnow for pointers to the literature and other
help.  None of them has read the article, so none of them are to blame
for any errors here and, though we consult each other carefully about
various aspects of our course designs, my UR~theory colleagues (the
plural pronoun prose style notwithstanding) are in no way to blame
for the opinions/projects described here.

}

\bibliographystyle{alpha}

{\singlespacing
\small

\bibliography{gry}

}

\end{document}